\documentclass{ar-1col}
\usepackage[numbers]{natbib}
\usepackage{url,amsmath}
\jname{Annu. Rev. Cond. Matt. Phys.}
\jvol{8}
\jyear{2016}
\doi{10.1146/((please add article doi))}

\let\a=\alpha \let\b=\beta \let\g=\gamma \let\d=\delta
 \let\z=\zeta \let\h=\eta 
\let\l=\lambda \let\m=\mu   
 \let\t=\tau \let\f=\varphi 
   \let\G=\Gamma
\let\D=\Delta \let\Th=\Theta \let\X=\Xi 
\let\Si=\Sigma   
 \let\r=\rho \let\th=\theta \let\io=\infty

 \def\VV{{\cal V}}
\def\FF{{\cal F}} 
 \def\BB{{\cal B}}
\def\RR{{\cal R}}  
\def\DD{{\cal D}}\def\AA{{\cal A}}

\newcommand{\wh}{\widehat}

\newcommand{\barr}{\begin{eqnarray}}
\newcommand{\earr}{\end{eqnarray}}
\newcommand{\beq}{\begin{equation}}
\newcommand{\eeq}{\end{equation}}
\newcommand{\be}{\begin{equation}}
\newcommand{\ee}{\end{equation}}

\newcommand{\de}{\mathrm{d}}
\newcommand{\eee}{\mathrm{e}}

\newcommand{\gv}[1]{\ensuremath{\mbox{\boldmath$ #1 $}}} 
 
\newcommand{\grad}[1]{\gv{\nabla} #1} 
\let\baraccent=\= 
\renewcommand{\=}[1]{\stackrel{#1}{=}} 

{\left\lbrace\begin{array}{@{}l@{}}}%
{\end{array}\right.}

\newcommand{\R}{\mathsf{R}}

\begin{document}

\markboth{CKPUZ}{Glass and Jamming Transitions}

\title{Glass and Jamming Transitions: From Exact Results to Finite-Dimensional Descriptions}

\author{Patrick Charbonneau,$^1$ Jorge Kurchan,$^2$ Giorgio Parisi,$^3$ Pierfrancesco Urbani, $^4$ and Francesco Zamponi$^5$
\affil{$^1$Department of Chemistry, Duke University, Durham, NC 27701, USA.}
\affil{$^2$LPS, \'Ecole Normale Sup\'erieure, UMR 8550 CNRS, 24 Rue Lhomond, 75005 Paris, France.}
\affil{$^3$Dipartimento di Fisica,
Sapienza Universit\`a di Roma,
INFN, Sezione di Roma I, IPFC -- CNR,
Piazzale Aldo Moro 2, I-00185 Roma, Italy.}
\affil{$^4$ Institut de Physique Th\'eorique, Université Paris Saclay, CEA, CNRS, F-91191 Gif-sur-Yvette, France.}
\affil{$^5$ Laboratoire de Physique Th\'eorique, ENS \& PSL Univeristy, UPMC \& Sorbonne Universit\'es, 
UMR 8549 CNRS, 75005 Paris, France.}}

\begin{abstract}
Despite decades of work, gaining a first-principle understanding of amorphous materials remains an extremely challenging problem. However, recent theoretical breakthroughs have led to the formulation of an exact solution of a microscopic model in the mean-field limit of infinite spatial dimension, and numerical simulations have remarkably confirmed the dimensional robustness of some of the predictions. This review describes these latest advances. More specifically, we consider the dynamical and thermodynamic descriptions of hard spheres around the dynamical, Gardner and jamming transitions. Comparing mean-field predictions with the finite-dimensional simulations, we identify robust aspects of the theory and uncover its more sensitive features. We conclude with a brief overview of ongoing research.
\end{abstract}

\maketitle

\tableofcontents

\section{INTRODUCTION}
Although amorphous materials, such as grains, foams and glasses, are ubiquitous, their theoretical description remains rather rudimentary. Compared with ordered solids, which are central to any solid state textbook, the contrast could hardly be more glaring. The glass transition is an entirely different phenomenon than the first-order transition into a crystal.  A glass is not the thermodynamic ground state, so the ``glass transition'' occurs as a liquid falls out of equilibrium and becomes rigid.
The rigidity of glasses hence relies on altogether different principles than for crystals, and their materials properties bear that imprint.

Developing a theoretical description of amorphous materials is extremely challenging. Conventional paradigms for describing condensed matter rely on perturbative treatments around either the low-density, ideal gas limit (for moderately dense gases and liquids),
or an ideal lattice (for crystals). For amorphous materials, however, both strategies fail badly. Because these materials interact strongly, low-density starting points are unreliable, and equilibrium particle positions are not forthcoming. A natural controlled reference system around which to make a small-parameter expansion is thus missing. Understanding the rich non-equilibrium behavior of these materials, including the glass and the jamming transitions, has instead been left to uncontrolled treatments, differently balancing rigor and guesswork. Constructing a first-principle theory of amorphous materials thus remains one of the major challenges of theoretical condensed matter physics. 

Amorphous materials are not alone without a natural reference for building a perturbative description~\cite{1N}. 
A similar situation is encountered in
liquids~\cite{WRF87,FP99}, strongly coupled electrons~\cite{GKKR96}, 
atomic physics~\cite{atomic}, and
gauge field theory \cite{DPS79}.
For all these problems, dimensional expansion has been developed as an alternate line of attack. By solving the problem in infinite-dimensional space, $d\rightarrow\infty$, and treating $1/d$ as a small parameter, one may hope to recover--to a good approximation--the behavior of physical systems in $d=3$. Such a program for glasses was first proposed by Kirkpatrick, Thirumalai and Wolynes in the late 1980s~\cite{KW87}, as part of a larger effort to construct the ``Random First Order Transition'' (RFOT) theory of the glass transition~\cite{KW87b,KT87,KT87b,KT88,KT89,KTW89}.
Yet advances have been a long time coming, because the necessary tools to first solve the $d=\infty$ problem were not readily available. It is only a decade ago that some of us started attacking the problem head first for the simplest model glass former, hard spheres~\cite{PZ06a,PZ10}.

The large-$d$ approach may be a successful
calculation device in various fields of physics, but for glasses it plays an additional--perhaps even more important--role. It helps clearly define entities that are otherwise only phenomenological.
Consider, for instance, the ``configurational entropy", i.e., the ``entropy of glassy states"~\cite{Ka48}. In order to define this quantity properly,
one needs to count the logarithm of the number of metastable glassy states. However, it is not
possible to have an exponential number of states that are absolutely stable in a finite-$d$ system~\cite{BB04}.
One thus defines complexity as
the logarithm of the number of states with a lifetime larger than an (arbitrarily chosen) $t^*$.
The large-$d$ approach, by contrast, allows the precise enumeration of states that are stable in the limit $d\rightarrow\infty$, thus providing a clear-cut definition. 
 A related question is the definition of {\em activated processes}, which are notably responsible 
for blurring the dynamical glass transition. It seems mathematically clear and physically plausible that
activated processes may be identified as those that take an exponentially long time, $t \sim e^{\mathcal{O}(d)}$, to complete. This
definition gives a hint of the (nonperturbative) techniques needed to study such processes.
It may even be that the entire RFOT scenario is but a metaphor based on
entities that only truly exist in the limit of large dimensions.

The essential counterpart to the large $d$ computation is to assess the robustness of the physical phenomena under changing $d$. There would be but limited relevance to a $d=\infty$ solution for processes that acutely depend on spatial details. Consider, for instance, ordered packings of hard spheres in $\R^d$. Asymptotic packing bounds for $d\rightarrow\infty$~\cite{CS99} have little to say about the singularly dense triangular lattice in $d=2$, root lattice in $d=8$~\cite{Vi16}, or Leech lattice in $d=24$~\cite{CKMRV16}. Does something similar occur for amorphous materials? Computational answers to this question have been emerging from various directions over the last decade~\cite{SDST06,ER09,MFC09,MCFC09,CIMM10,CIPZ11,CCT12,CCT13,SKDS13}. It has now become clear that many features of these materials are, at least qualitatively and sometimes even quantitatively, remarkably independent of $d$. 

The main purpose of this review is to describe the astounding theoretical and numerical advances triggered by these realizations. For the sake of concision and in order to be relatively self-contained, we here only consider hard-sphere glass formers. Note, however, that this choice is largely done without loss of generality, as we argue in the conclusion.

\subsection{Setup}
We consider the behavior of a system of $N$ hard spheres of diameter $\DD$ within a box of volume $\VV$, in $d$ spatial dimensions. We denote $x_i$ the $d$-dimensional vector that encodes the coordinates of sphere $i$, with $i=1,\ldots, N$,
and $x_i^\m$ its coordinates with $\m=1\cdots d$.
The singular nature of the hard-core interaction results in temperature playing but a trivial scaling role. Hence, the only equilibrium control parameter is the packing fraction, $\f$, i.e., the fraction of $\VV$ occupied by spheres. For the number density of spheres $\r=N/\VV$, we thus have $\f=\r \VV_\DD$, where $\VV_\DD$ is the $d$-dimensional volume of a sphere of diameter $\DD$. Throughout the text we also rescaled version of physical quantities, $\widehat{\cdot}$, such that their numerical values remain of order unity in the $d\rightarrow\infty$ limit, e.g., $\wh\f=2^d\f/d$.

In the following, Section~\ref{sec:dynamics} presents the hard sphere liquid dynamics and compares the exact solution with mode-coupling theory. Section~\ref{sec:compression} details what happens to an out-of-equilibrium amorphous solid upon a quasi-static compression, and the jamming endpoint to these compressions is separately discussed in Section~\ref{sec:J}. Section~\ref{sec:eq_ideal_glass} presents the results of various out-of-equilibrium processes. Each section first describes the conceptual framework and the theoretical results from the exact $d\rightarrow\infty$ solution, 
and then compares these with experimental and numerical results for finite-$d$ systems.
Note, however, that when the $d\to\io$ solution straightforwardly reproduces well-known results from other approaches  that have been extensively reviewed elsewhere, 
we limit ourselves to a minimal discussion of the low-$d$ results.

\section{EQUILIBRIUM DYNAMICS AND MODE-COUPLING THEORY}
\label{sec:dynamics}

Consider the following experiment: take at random a valid configuration of hard spheres at density $\f$,
randomly assign each sphere a velocity from the Maxwell distribution, and evolve the Newtonian dynamics.
At very low $\f$, dynamics is mostly ballistic, but the rare collisions eventually give rise to a diffusive regime. The system is then nearly ideal gas-like. Increasing $\f$ makes dynamics more liquid-like, as particles move a distance no further than $\DD$ before their travel direction gets randomized by  collisions and their diffusive behavior ensues.
Further increasing $\f$ from this regime is the interest of this review.


If crystallization is successfully avoided--see \cite{ZSWLSSO14,SDST06,MFC09,MCFC09} for practical ways of achieving this, 
the liquid dynamics grows increasingly sluggish with $\f$. Sphere diffusivity decays and, correspondingly, the structural relaxation time, $\t_\a$, grows so much that the system remains structurally similar to its initial configuration over experimentally accessible timescales. This loss of ergodicity defines the experimental glass transition. Why does diffusive dynamics become so slow and eventually freeze?
In the $d\to\io$ limit we can answer this question precisely because the equilibrium dynamics of hard spheres has been solved \cite{KMZ16}. In this section, we review this result and comment on its connection with other theoretical descriptions. We then discuss how it relates to the phenomenology of finite-$d$ systems.

\subsection{Mean-field dynamical equations}

In order to solve the liquid dynamics, it is convenient to substitute Newtonian by Langevin dynamics and to momentarily relax the hard-core constraint. We thus assume that spheres interact through a pair potential $V(r) = \overline V[ d(r-\DD) ]$, where $\overline V(h)$ remains finite for $d\to\io$. The hard sphere limit corresponds
to $\overline V(h)=0$ for $h>0$ and $\overline V(h)=\io$ for $h<0$.
The system dynamics is then obtained by solving
\beq\label{eq:Langdyn}
m \ddot x_i(t)+\z \dot x_i(t)=-\grad_{x_i}H+\xi_i(t) \ ,
\eeq
where $m$ is the mass of the spheres, $H=\sum_{i<j}V(|x_i-x_j|)$ is the system energy, $\z$ is the friction (or drag) coefficient, and $\xi_i$ is white noise with zero mean and variance $\langle \xi_i^\mu(t)\xi_j^{\mu'}(t')\rangle=2T\z \delta_{ij}\delta_{\mu \mu'}\delta(t-t')$ with temperature $T=1/\b$.
We are interested in understanding the behavior of the dynamical response and correlation functions,
\beq
R(t,t') =\frac{2d}{\DD^2N }\sum_{i=1}^N \sum_{\mu=1}^{d}\frac{\delta x_i^\mu(t)}{\delta \xi_i^\mu(t')} \mbox{ and }\Delta(t,t')=  \frac{d}{\DD^2 N}\sum_{i=1}^N \left|x_i(t)-x_i(t')\right|^2 \,.
\eeq
Equilibrium dynamics is described by starting the Langevin process in an equilibrated configuration 
at density $\f$ and temperature $T$; in this case the solution of the dynamical equations satisfies both time translational invariance, i.e., $\D(t,t')=\D(t-t')$ and $R(t,t')=R(t-t')$, and the fluctuation-dissipation theorem, i.e., $R(t) = \b \th(t) \dot \D(t)$~\cite{CK93,Cu02}.

For $d\rightarrow\infty$, and in the thermodynamic limit $N\to\io$, $\D(t)$ obeys~\cite{KMZ16}
\begin{eqnarray}
\wh m \ddot \D(t) +
\wh \z \dot \D(t)=T-\b\int_0^t\de u\, M(t-u)\dot \D(u) \ ,
\label{eq:MCT}
\end{eqnarray}
where $\wh \z=\DD^2\z/(2d^2)$, $\wh m =\DD^2m/(2d^2)$, and the memory kernel $M(t)$ is the solution of the self-consistent equations
 \begin{equation}
 \label{eq:Mselfcons}
 \begin{split}
& \wh m \ddot y(t) + \wh \z \dot y(t)= T - \overline V'[y(t)]-\b \int_0^t\de u M(t-u)\dot y(u)+\X(t) \\
\mbox{ and } & M(t-t')=\frac{\wh \f}{2}\int \de y_0\eee^{y_0 -\b \overline V(y_0)} \langle F[y(t)]F[y(t')]\rangle_\X \ ,
 \end{split}
 \end{equation}
 where $y(t=0)\equiv y_0$, $\langle\X(t)\X(t') \rangle=2\wh \z T\d(t-t')+M(t-t')$, and $F(y)=-\overline V'(y)$ is a scaled interaction force.
 
The complex problem of $N$ interacting particles in $d$ dimensions 
has thus been reduced to a simple one-dimensional problem: an effective degree of freedom moving in an effective potential 
$\overline V(h) - T y$
in presence of a colored noise, whose memory kernel $M(t)$ is determined self-consistently as the average of a force-force correlation.
 Remarkably, these equations are akin to those from the generalized schematic mode-coupling theory (MCT) \cite{Bouchaud,Go08}. 
The key difference is that within the MCT approximation, the memory kernel is a simple function of $\Delta(t)$ alone (essentially, $M \sim \D^2$), 
while the exact expression for $M(t)$ in $d\rightarrow\infty$ is given by Eq.~\eqref{eq:Mselfcons} as an implicit functional of $\D(t)$.
Because the critical scaling of the dynamical equations is insensitive to the precise form of $M(t)$, however, many qualitative features of standard MCT persist in that case as well.

Recall that the above result was derived for a soft potential $V(r)$ and for a Langevin dynamics with friction and noise, but one
 can safely take the hard sphere limit of $V(r)$ and the limit $\z\to 0$ at which the noise disappears and the Langevin dynamics becomes Newtonian.
The latter limit, however, should be taken after taking the limits $N\to \io$ and $d\to\io$. Newtonian dynamics is
then described by a self-consistent effective Langevin dynamics for which noise is generated by interactions. 
 
 \subsection{Dynamical transition}
\begin{figure}[h]
\includegraphics[width=\columnwidth]{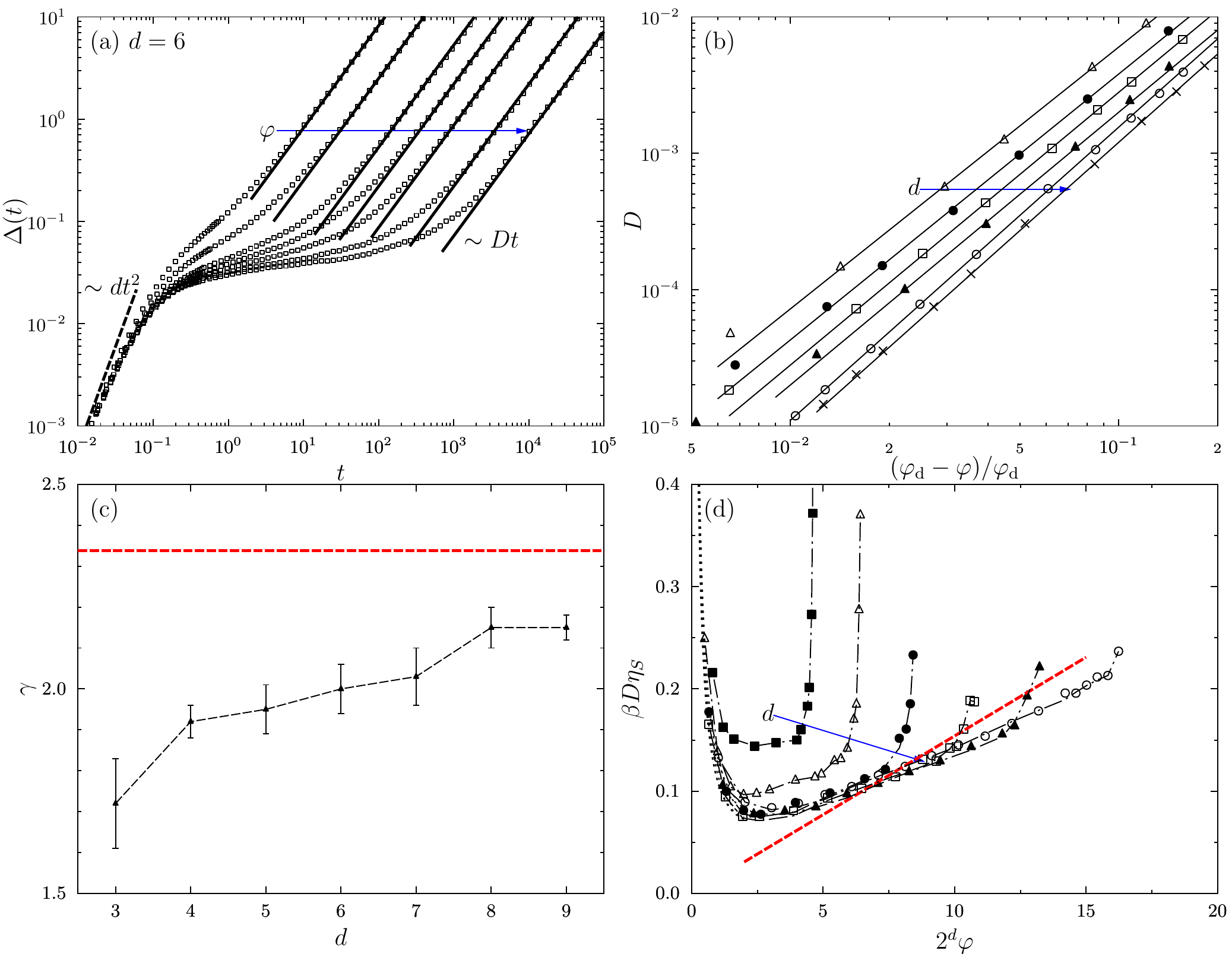}
\caption{
(a)  Time evolution of the mean-square displacement $\D(t)$ in the liquid phase for $d=6$~\cite{CIPZ12}. 
As density increases, the caging regime that separates the short-time ballistic, $\sim dt^2$ and the long-time diffusive regimes, $\sim Dt$, grows longer.
(b) The decay of the diffusivity becomes more power-law like as dimension increases, in the dynamical regime accessible in simulations~\cite{CIPZ12}.
(c) The critical MCT exponent $\gamma$ extracted from simulations steadily approaches the prediction for $d\rightarrow\infty$ (red line) as $d$ increases~\cite{CJPZ14}.
(d) The relationship between the transport coefficients $D$ and $\eta_\mathrm{S}$ has a non-trivial density dependence that qualitatively follows the high-$d$ prediction (red line)~\cite{CCJPZ13}.
}
\label{fig:MSD}
\end{figure}
The order parameter for the dynamical transition is the long time limit of the dynamical mean-square displacement, $\Delta(t)$.
At low $\f$, the short-time ballistic behavior $\Delta(t)\sim t^2$ is promptly followed by a diffusive regime $\Delta(t)\sim \wh D t$.
The dynamical equation directly gives the (rescaled) diffusion coefficient~(\ref{eq:MCT}), and a similar analysis gives the shear viscosity~\cite{KMZ16}, 
\beq \label{eq:viscosity}
 \wh D= \frac{2 d^2}{\DD^2} D =  \frac{T}{\wh \g +\b\int_{0}^\infty\de u\, M(u)}\mbox{ and } \wh \h_S = 
 \frac{2^d\, \VV_\DD}{d^2}  \h_S =
 \b \wh\f \int_0^\io \de u\, M(u)\;,
\eeq
hence the two transport coefficients obey
\beq\label{eq:SER}
\wh D= \frac{T}{\wh \g +\wh\h_S/\wh\f }\:.
\eeq
One of the hallmarks of sluggish dynamics is that upon increasing $\wh\f$, the ballistic and diffusive regimes grow separated by a plateau at $\Delta_{\mathrm{EA}}$, as in Fig.~\ref{fig:MSD}.
The dynamical equations indicate that for $d\to\infty$ the length of the plateau diverges at the dynamical transition $\wh \f_\mathrm{d}$.
The long-time limit of the mean-square displacement at that density is thus finite, i.e., $\lim_{t\rightarrow\infty}\Delta(t)=\Delta_{\mathrm{EA}}<\infty$.
The system then cannot equilibrate and remains stuck in a glass state. 
Correspondingly, $M(t)$ also develops a plateau, $\lim_{t\rightarrow\infty} M(t)=M_{\mathrm{EA}}>0$, indicating that memory of the initial condition
persists at all times. As a consequence,
$\int_{0}^\infty\de u\, M(u) \to \io$, diffusivity vanishes and viscosity diverges.
Yet $\wh D \wh \h_S/T \sim \wh\f$ remains finite, which is reminiscent of the Stokes-Einstein relation.

Upon approaching the dynamical transition, a liquid keeps an increasingly long memory of the initial configuration and remains in its vicinity. Both $\wh \f_\mathrm{d}$  and a measure of that proximity, the plateau height $\D_\mathrm{EA}$, 
can be obtained from the long time limit of the dynamical equations (\ref{eq:Mselfcons}) as~\cite{PZ10, KPZ12}
\beq\label{eq:1RSB_f}
\begin{split}
\Delta_{\mathrm{EA}}&=\textrm{argmax}_{\D}\FF_1(\D)  \mbox{ \ \ \ \ \ and \ \ \  \ \     }
\wh \f_\mathrm{d}=\frac{1}{\FF_1(\Delta_{\mathrm{EA}})} \ ,\\
\mbox{where }\FF_1(\D)&=-\D \int_{-\infty}^\infty\de h\, \eee^{h} \frac{\de }{\de \D} \left[\Th\left[\frac{h+\D/2}{\sqrt{2\D}}\right]\ln \Th\left[\frac{h+\D/2}{\sqrt{2\D}}\right]\right] \ ,
\end{split}
\eeq
with $\Th(x)=[1+\textrm{erf}(x)]/2$. We will see in Section~\ref{sec:compression} that these results can also be obtained using a completely static approach.
 
\subsection{Dynamical critical exponents and susceptibility}

As we mentioned above, the dynamics exhibits a critical scaling around $\wh\f_{\rm d}$ akin to standard MCT~\cite{Go08}.
In particular, close to $\wh \f_\mathrm{d}$, $\D(t)$ scales differently upon approaching and leaving the plateau
 \begin{eqnarray}
 \Delta(t)\simeq \D_{\mathrm{EA}}- \AA t^{-a} \mbox{ and } \Delta(t)\simeq \D_{\mathrm{EA}}+ \BB t^b \ ,
 \end{eqnarray}
with critical exponents that are related through a non-universal exponent parameter \cite{Go08}
 \beq
 \l=\frac{\G^2(1-a)}{\G(1-2a)}=\frac{\G^2(1+b)}{\G(1+2b)} \ .
 \label{eq:lambda}
 \eeq
For hard spheres in $d\to \infty$, we find $\l=0.70698...$~\cite{KPUZ13}.
The dynamical equations additionally provide the power-law divergence of the viscosity and $\tau_\a$ (and decay of $\widehat{D}$), $\h_S \sim \t_\a\sim \widehat{D}^{-1}\sim|\f-\f_\mathrm{d}|^{-\g}$, where $\g=1/(2a)+1/(2b)$. 

At a standard phase transition, one commonly detects a critical point by considering the correlation function of the order parameter and its integral over $\mathcal{V}$.
By analogy, in the case of the dynamical transition one can study the dynamical susceptibility 
\beq
\chi_4(t)=N\left(\overline{\Delta^2(t)}-\overline{\Delta(t)}^2\right) \ ,
\eeq
where the overline denotes 
averaging a quantity over both the system thermal history, $\langle \cdot \rangle$, and initial configurations $\{x_i(0)\}$. 
$\chi_4(t)$ encodes the fluctuations of dynamical correlators, here represented by the mean square displacement.

In the glass phase, the long-time limit of this susceptibility goes to a constant, i.e., $\lim_{t\rightarrow\infty}\chi_4(t)=\chi$, that diverges upon approaching the dynamical transition from the glass phase ($\wh \f \to\wh \f_\mathrm{d}^+$) as $\chi\sim |\wh \f-\wh \f_\mathrm{d}|^{-1/2}$. 
Approaching the dynamical point from the liquid side instead, $\chi_4(t)$ peaks at $t\sim \t_\a$, with $\chi = \chi_4(\t_\a)$
similarly diverging~\cite{KT88,FP00,DFGP02,BBMR06,BBBKMR07a,BBBKMR07b,BBBCEHLP05,BBBCS11}.
In addition, a dynamical correlation length associated with $\chi_4$ diverges as $\xi_{\mathrm{d}}\sim |\wh\f-\wh \f_\mathrm{d}|^{-1/4}$ near the transition~\cite{FPRR11,FJPUZ12,FJPUZ13}.
 
\subsection{Liquid dynamics in finite dimensions}\label{sec:no_dyn}
 
The dynamical transition at $\wh \f_\mathrm{d}$ found in the $d\to\infty$ limit is quite fragile. It does not formally exist in any finite dimension~\cite{BNT15}. 
Activated processes sidestep the dynamical arrest, hence all configurations have a finite lifetime~\cite{KTW89}.
Despite this difficulty, we can nevertheless probe what aspects of the $d\rightarrow\infty$ transition have a lower-dimensional echo. 

Because many physical features of the exact dynamical solution qualitatively coincide with those of standard MCT (the latter preceding the former by more than 30 years), substantial efforts have already been devoted to evaluating some of the theoretical predictions~\cite{Go08}. 
The two-step behavior of $\D(t)$ (Fig.~\ref{fig:MSD}a), the scaling relations between the critical dynamical exponents, 
and the growing dynamical susceptibility that accompanies the lengthening plateau in $\D(t)$,
have indeed been carefully documented in $d=3$~\cite{BBBCS11}.

Because activated processes are expected to disappear exponentially quickly when dimension increases, however, one might also hope for quantitative aspects of the theory to be validated by numerical simulations in higher $d$. In this respect, a few results are noteworthy. (i) The dynamical range over which a power-law scaling of $\tau_\alpha$ is observed steadily increases with $d$ (see Fig.~\ref{fig:MSD}b). From barely larger than a decade in $d=3$, it grows to nearly three decades (before exhausting computational resources) in $d=8$~\cite{CIPZ12}. (ii) If one carefully avoids the activated dynamical regime, the exponent parameter $\l$ that appears in Eq.~(\ref{eq:lambda}) tends toward the $d\rightarrow\infty$ prediction as $d$ increases~\cite{CJPZ14} (Fig.~\ref{fig:MSD}c). 
(iii) The difference between standard MCT and the exact $d\rightarrow\io$ results for $M(t)$ gives rise to diverging asymptotic predictions for $\wh \f_\mathrm{d}$~\cite{IM10,SS10}. While numerical simulations in $d=2$ to 13 are more consistent with the exact solution~\cite{CIPZ11}, even in $d=12$ the asymptotic scaling regime remains distant. Efforts at systematically computing finite-$d$ corrections to the exact theory have been made~\cite{PZ10}, but the subject remains an open area of research~\cite{MZ16}. (iv) The scaling $\tau_\alpha\sim \wh D^{-1}$, often dubbed the Stokes-Einstein relation in liquids, is violated abruptly in finite-$d$ glass formers. As $d$ increases, however,  this violation gets pushed back to increasingly long timescales, revealing the linear correction to the Stokes-Einstein scaling predicted by Eq.~(\ref{eq:SER})~\cite{CIMM10,CCJPZ13,CCJPZ13,KMZ16} (Fig.~\ref{fig:MSD}d).

A key prediction of the exact solution that has yet to be convincingly tested is the critical scaling of $\chi_4$ and $\xi_{\mathrm{d}}$ --
see Ref.~\cite{CIM15} for a test in a different model. Because increasing $d$ enlarges the range of power-law scaling of $\tau_\alpha$, 
the divergence of $\chi_4$ should then also become sharper. The significant computational efforts needed to extract this quantity in $d>3$ has thus far discouraged attempts in that direction, but its time will certainly come. 



\subsection{Exploring dynamics through equilibrium calculations}

The dynamical equations that describe the equilibrium liquid in the limit $d \rightarrow \infty$ reveal that, upon approaching
a critical density, the relaxation timescale diverges. Equilibration is impossible beyond that point. An equilibrium liquid that is ``crunched'' (quickly compressed)
to such densities does not, however, immediately acquire an infinite relaxation time, but {\em ages} instead; as time passes, $\Delta$ relaxes
ever more slowly. Hence, $\Delta(t,t_w)$ becomes a function of both the (waiting) time $t_w$ elapsed since the crunch
and of the time difference $t-t_w$~\cite{CK93,KMZ17}.
Given that equilibrium is not achieved over reasonable times in this regime, equilibrium calculations may seem
of limited interest. By modifying the set of configurations considered, i.e., the measure, however, one may nonetheless hope to extract
dynamical information. And because they are often technically simpler--using the replica and cavity methods~\cite{MPV87}--than their dynamical counterpart, a panoply of methods have been developed. Results for various equilibrium methods are presented in the following sections, but before proceeding
this brief overview aims to guide disoriented readers through some of the relevant spin-glass literature, where these ideas were first developed. 
Note that the precise dynamical justification, 
varies from scheme to scheme. In most cases, however, such justification is only known for a normal glass, and it remains unclear how to interpret equilibrium calculations in a marginal glass. 

{\em Counting metastable states:} Bray and Moore~\cite{BM80}
first showed that the number of metastable states  (as defined by the Thouless-Anderson-Palmer equations, assumed
by Bray and Moore to be relevant for dynamics) is exponential. This computation can also be implemented from the dynamical equations~\cite{BK01}. 

{\em Dynamics starting from various equilibrium measures:}
The dynamical equations can be modified to describe exactly the evolution
of an equilibrium configuration at a different temperature or density~\cite{BBM96}.
In a normal glass phase, the long time evolution can be described by computing the probability
distribution of a configuration kept at fixed distance away from a reference one~\cite{FP95};
this construction -- reviewed in Sec.~\ref{sec:compression} -- accesses certain
quantities associated with a full dynamical computation~\cite{BBM96}. 
Once can also extend it to an entire chain of
copies, which is conjectured to mimic slow dynamics in its full generality, including in the marginal glass phase~\cite{FP13,FPU15}.

{\em Threshold level and marginality:}
In a normal glass, computing the level at which states become marginally stable
-- as measured by their diverging four-point correlations--is justified from the dynamical solution. Without any special assumption, this analysis finds that
aging takes place at free energies above which the landscape connects and below which it disconnects~\cite{CK93}.
This threshold level corresponds to the neighborhood
of marginal states. We review this construction in Sec.~\ref{sec:eq_ideal_glass}.

{\em  Effective temperatures via Edwards-like assumptions:}
Another feature that appeared in the dynamical solution, and that validates a quasi-equilibrium approach, is
that during aging the system samples states with a fixed energy {\em uniformly}.
This assumption, which is equivalent to that proposed
by Edwards for granular matter~\cite{EO89}, is surprisingly exact within
the dynamical solution~\cite{Ku01,CR00,FV00}.
We review this construction in Sec.~\ref{sec:eq_ideal_glass}.

{\em Parisi matrices as generating functions of (certain) dynamical diagrams:}
A recent application of replicas exploits the mathematical correspondence
between the diagrams generated by Parisi matrices, which can be efficiently evaluated,
 and those used to compute some dynamic quantities~\cite{CFLPRR12,PR12},
such as the exponents for time relaxation within and away from a state.
This correspondence is purely mathematical -- it implies no physical assumption.


\section{FOLLOWING GLASSES UNDER SLOW COMPRESSION} 
\label{sec:compression}

As argued in Sec.~\ref{sec:dynamics}, the diffusion constant strictly vanishes beyond the dynamical transition for $d\rightarrow\infty$, because the timescale for activated processes grows exponentially quickly with $d$. The system thus remains trapped for an extremely long (infinite) time into a restricted region
of phase space, which defines a metastable state.
Insights into this regime can then be obtained from statistical mechanics alone--without solving the dynamical equations--using tools developed by Franz and Parisi~\cite{FP95,BFP97,RUYZ15, RU15}. In this section, we first formalize the notion of metastable states and present the exact solution for compressing equilibrium configurations at $\wh\f>\wh\f_\mathrm{d}$ for $d\rightarrow\infty$.  We then discuss how such configurations can be obtained in finite $d$, and compare simulated compressions with theoretical predictions.

\subsection{Metastable states}
\label{sec:DDr}
First, consider an equilibrium configuration $Y$ at $\wh \f_\mathrm{g}>\wh \f_\mathrm{d}$, and a configuration $X(t)$ that is dynamically evolved following Eq.~(\ref{eq:Langdyn}) from the initial condition $X(0) = Y$.
Because diffusivity vanishes in this regime for $d\to\io$, i.e., $D=0$, the relative mean square distance between $X(t)$ and $Y$ remains finite, even in the long-time limit, i.e.,
\beq\label{eq:Drdef}
\lim_{t\rightarrow\infty}\D(X(t),Y) = \lim_{t\rightarrow\infty}\frac{d}{N\DD^2}\sum_i \left|x_i(t)-y_i \right|^2 =\D_\mathrm{r} \ .
\eeq  
Moreover, because $Y$ is sampled from the equilibrium distribution, the dynamics of $X(t)$ must be stationary, i.e.,
\beq
\D(X(t+\t),X(t))= \D(\t) \to_{\t \to\io}\D_\mathrm{r}\ ,
\hskip30pt \forall t  > 0\ .
\eeq

Next, consider first compressing (or decompressing) the initial configuration $X(0)$ to $\wh\f \neq \wh\f_{\rm g}$, and then evolving the system dynamics 
 at this new density. If the diffusion constant remains zero, then we expect 
Eq.~(\ref{eq:Drdef}) to hold, and 
$X(0)$, $X(t)$ and $X(t+\t)$ to always be close to one another.
Although the overall dynamics for this protocol is not stationary in general, it nonetheless becomes so at long times, i.e.,
\beq
\lim_{t \to \io} \D(X(t+\t),X(t)) = \D(\t) \to_{\t \to\io} \D \ .
\eeq
In other words, $\D_\mathrm{r}$ is the relative long-time displacement between $X$ and $Y$, while $\D$ is the relative displacement
between configurations $X$ at very different times. For $X$ and $Y$ at the same density $\D=\D_\mathrm{r}$, but otherwise the two quantities differ.

In finite $d$ the configuration $X$ is not expected to remain trapped around $Y$ for an infinite time. This residence time, however,  can be made sufficiently long for a clear separation of timescales between sampling a glass state and leaving that state. 
The region initially visited by $X$ around $Y$ is thus only a metastable state. 

\subsection{The Franz-Parisi restricted free energy and the order parameters}

We now wish to translate the dynamical formulation of Sec.~\ref{sec:DDr} into a purely thermodynamic description.
Consider again an equilibrium configuration $Y$ at $\wh \f_\mathrm{g}>\wh \f_\mathrm{d}$ that is evolved to a metastable state visited by $X$ at a packing fraction $\wh \f\neq\wh\f_\mathrm{g}$. From the discussion in Sec.~\ref{sec:DDr}, we know that in the long-time limit the distance between $X$ and $Y$ is $\D_\mathrm{r}$.
The minimal assumption about $X$ is that it is sampled by
the Boltzmann distribution at $\wh\f$, under the constraint that it maintains its average distance
to $Y$. We thus get
\begin{eqnarray}
 P(X, \wh \f|Y, \wh \f_\mathrm{g})&=\frac{\eee^{-\b H[X; \wh \f]}}{Z[\D_\mathrm{r},\wh\f | Y, \wh \f_\mathrm{g}]}\delta(\D_\mathrm{r}-\D(X,Y)) \ , \\
\mbox{with } Z[\D_\mathrm{r},\wh\f | Y,\wh \f_\mathrm{g}]&=\int \de X \eee^{-\b H[X;\wh \f]}\delta(\D_\mathrm{r}-\D(X,Y)) 
 \end{eqnarray}
and $\D_\mathrm{r}$ left unspecified for now. (Recall that for hard spheres these quantities do not actually depend on $\b$.)
We can then define the generalized free energy of a metastable
state selected by $Y$
 \beq
 \mathsf{f}(\D_\mathrm{r}, \wh\f |Y,\wh \f_\mathrm{g})=-\frac{1}{N\b}\ln Z[\D_\mathrm{r},\wh\f|Y,\wh \f_\mathrm{g}].
 \label{eq:f_single_sample}
 \eeq
This quantity is a random variable that depends on the initial configuration $Y$; because different $Y$ select different metastable states the free energy fluctuates. 
The free energy, however, self-averages. Hence, in the thermodynamic limit it concentrates on  its mean value and it suffices to compute the average of
 $ \mathsf{f}(\D_\mathrm{r}, \wh\f |Y,\wh \f_\mathrm{g})$ over $Y$. We then obtain the Franz-Parisi free energy (or potential)
\begin{eqnarray}
 V_{\mathrm{FP}}(\D_\mathrm{r}, \wh \f | \wh \f_\mathrm{g})=\overline{\mathsf{f}(\D_\mathrm{r},\wh\f | Y,\wh \f_\mathrm{g})}^Y=-\frac{1}{\b N} \int \frac{\de Y}{Z[\wh \f_\mathrm{g}]}\eee^{-\b H[Y;\wh \f_\mathrm{g}]}\ln Z[\D_\mathrm{r},\wh\f|Y,\wh \f_\mathrm{g}] \label{FP_potential_SF} \ ,
 \end{eqnarray}
where $Z[\wh \f_\mathrm{g}]=\int \de Y \eee^{-\b H[Y;\wh \f_\mathrm{g}]}$ is the equilibrium partition function at $\wh\f_\mathrm{g}$.
 
For $d\rightarrow\infty$,  the potential $V_{\mathrm{FP}}(\D_\mathrm{r}, \wh \f | \wh \f_\mathrm{g})$ 
can be computed explicitly though the replica method~\cite{RUYZ15}, and expressed as a function of the two order parameters,
$\D_\mathrm{r}$ and $\D$, introduced in Sec.~\ref{sec:DDr}.
Note that Eq.~(\ref{FP_potential_SF}) explicitly  depends on $\D_\mathrm{r}$, but $\D$ only appears upon unfolding the logarithm of the partition function by the replica
method. Its value is then chosen so as to minimize the free energy.
The value of $\D_\mathrm{r}$, by contrast, is obtained from the Franz-Parisi potential and thus has a richer physical interpretation. If $V_{\mathrm{FP}}(\D_\mathrm{r}, \wh \f | \wh \f_\mathrm{g})$  displays a local minimum at a finite
$\D_\mathrm{r}$, then the probability of finding $X$ a distance $\D_\mathrm{r}$ away from $Y$ is high, and the metastable state is characterized by this $\D_\mathrm{r}$. If the potential shows no local minimum at finite $\D_\mathrm{r}$, then $Y$ is unable to trap $X$ at long times, and the system is liquid.

\begin{figure}
\includegraphics[width=\textwidth]{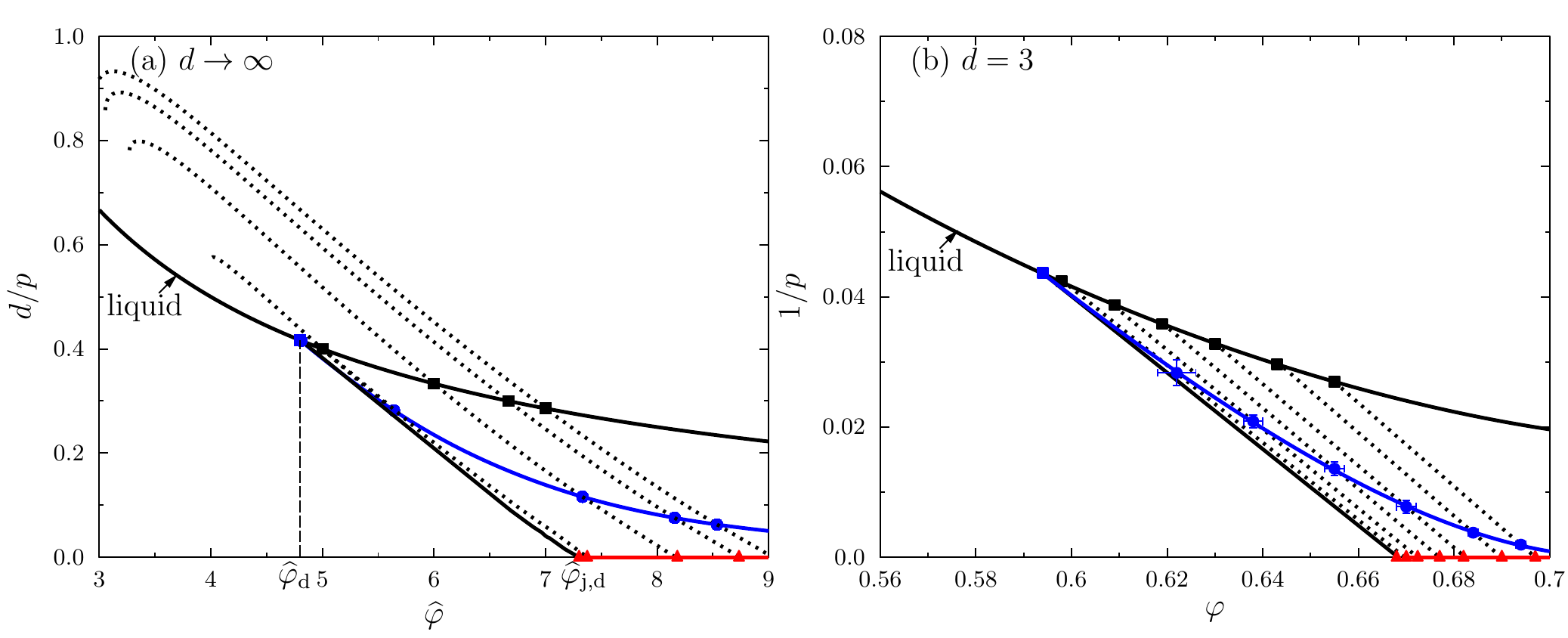}
\caption{(a) Phase diagram summarizing the exact results for hard spheres in the limit $d\rightarrow\infty$. Dashed lines correspond to the equations of state of glasses prepared at different initial packing fractions $\wh \f_\mathrm{g}$ (black squares). The limit case is the solid line at $\wh\f_{\mathrm{d}}$ (blue square). For $\wh \f_\mathrm{g}>\wh\f_{\mathrm{d}}$, decompression results in hysteresis, and upon compression a glass undergoes a Gardner transition (blue dot) before jamming (red triangle)\cite{RU15}. (b) Corresponding phase diagram for $d=3$ hard spheres~\cite{BCJPSZ15}.}
\label{fig:PD_SF}
\end{figure}

From this construction, we can determine the phase diagram of the metastable states. For each $\wh\f_\mathrm{g}$, 
this computation describes a set of equivalent
states that all have the same properties in the thermodynamic limit. For fixed $\wh\f_\mathrm{g}$, we can also follow the adiabatic
evolution of these states at different $\wh\f$, and obtain their pressure $p(\wh\f | \wh\f_\mathrm{g})$ from the derivative
of Eq.~(\ref{FP_potential_SF}) with respect to $\wh\f$~\cite{RUYZ15}. The resulting curve, $p(\wh\f | \wh\f_\mathrm{g})$, is the glass equation of state for a given $\wh\f_\mathrm{g}$ (see examples in Fig.~\ref{fig:PD_SF}). An especially remarkable feature of these curves is their hysteresis under decompression. A glass state can be followed up to a spinodal point, whereat it melts. Note that denser $\wh\f_\mathrm{g}$ display more extended hysteresis branches.


\subsection{Equilibrium Glasses}
In the above treatment, the case $\wh \f=\wh\f_\mathrm{g}$ corresponds to equilibrium glasses. 
As expected, we then obtain $\D=\D_\mathrm{r}=\D_\mathrm{EA}$. In addition, the pressure $p(\wh\f_\mathrm{g} | \wh\f_\mathrm{g})$ along this line is exactly the analytic continuation of the equilibrium liquid pressure beyond $\wh\f_\mathrm{d}$
(black line in Fig.~\ref{fig:PD_SF}a).
We thus obtain the following interpretation for the continuation of the liquid line in the regime $\wh \f>\wh\f_\mathrm{d}$: the system is frozen, with $D=0$, and the liquid phase is replaced
by a collection of glassy metastable states, each with the same pressure as the equilibrium liquid would have at that density.
In other words, while the system dynamics arrests completely at $\wh\f_\mathrm{d}$, its thermodynamics remains perfectly smooth and analytic.

The Franz-Parisi potential, by going beyond standard thermodynamics, also gives access to features of the dynamical transition. The transition, for instance, occurs at the smallest density for which a minimum at $\D_\mathrm{r}=\D_\mathrm{EA}$ exists. For $\wh\f_\mathrm{g} < \wh\f_\mathrm{d}$, the potential has no local minimum and the system is a liquid.
Pursuing this analysis provides the same results as the dynamical approach of Sec.~\ref{sec:dynamics} for the dynamical critical exponent $\l$ and
the critical scaling of $\chi_4$~\cite{CFLPRR12,KPUZ13}. 
There is a complete correspondence between the two approaches.

\subsection{Gardner transition}
\label{sec:SF_G}
An implicit assumption of the above discussion is that glass states prepared at $\wh \f_\mathrm{g}$ 
do not undergo a phase transition as $\wh \f$ increases.
A glass state can thus be described as an isolated set of configurations, with a typical mutual distance $\D$, a typical distance
from the reference configuration $\D_{\rm r}$, 
and a finite relaxation time. Each glass state is then an amorphous solid characterized by a well-defined shear modulus and non-linear elastic susceptibilities \cite{BU16}. 
Neglecting the Debye contribution to the vibrational spectrum, such solid also displays a gapped vibrational density of states \cite{FPUZ15,CCPPZ15}. In short, the behavior of this glass is characteristic of a normal solid. The normal glass scenario, however, is not the whole story. 
At a sufficiently high density $\wh \f_\mathrm{G}$--or pressure $p_\mathrm{G}$-- 
each glass state undergoes a Gardner phase transition. Beyond this $\wh\f_\mathrm{G}$ (the value depends on $\wh\f_\mathrm{g}$) the simple description in terms of two order parameters, $\D_\mathrm{r}$ and
$\D$, breaks down~\cite{KPUZ13,CKPUZ14NatComm,CKPUZ14JSTAT,RUYZ15}. 

The phase transition that ensues, which was first described in spin glasses by Gardner \cite{Ga85} and by Gross, Kanter and Sompolinsky \cite{GKS85}, provides a whole new paradigm for understanding glass properties.
At the Gardner transition, the region of phase space (or basin) to which a given glass state belongs develops a non-trivial internal structure of states described by the full-replica-symmetry-breaking solution of the partition function~\cite{MPV87, CKPUZ14NatComm, CKPUZ14JSTAT, RU15}. 
For $\wh\f_\mathrm{g}<\wh\f<\wh\f_\mathrm{G}(\wh\f_\mathrm{g})$, a glass state corresponds to a minimum in the free energy landscape, but at 
$\wh\f_\mathrm{G}(\wh\f_\mathrm{g})$ this minimum flattens, and for $\wh\f>\wh\f_\mathrm{G}(\wh\f_\mathrm{g})$ it transforms into a metabasin, i.e., a collection of metastable states with an ultrametric structure (Fig.~\ref{fig:Phase_space}). 


This nontrivial result is interesting for a number of reasons.
Consider first the impact of the Gardner transition on a glass state prepared at $\wh\f_\mathrm{g}$ and compressed up to $\wh \f$. 
In a normal glass, the time $\t_\b$ for equilibrating vibrations within the state (and thus for $\D(t)$ to reach its long time limit $\D$) is finite. Upon approaching $\wh\f_\mathrm{G}(\wh\f_\mathrm{g})$, however, 
this timescale diverges as a power law, $\t_\b\sim |\wh\f-\wh\f_\mathrm{G}|^{-a}$, 
with a non-universal critical exponent $a$ that depends on $\wh \f_\mathrm{g}$~\cite{PR12,RU15, BCJPSZ15,CJPRSZ15}.
This dynamical slowing down is a consequence of the second-order nature of the Gardner transition and of the associated soft modes. 
The emergence of these soft modes further translates into the divergence of the long-time limit of the dynamical susceptibility, $\lim_{t\rightarrow\infty}\chi_4(t)=\chi$, as $\chi\sim |\wh \f_\mathrm{G}-\wh \f|^{-1}$. This divergence is also related to the emergence of long-range correlations with a diverging correlation length $\xi_\mathrm{G}$~\cite{BCJPSZ15,CJPRSZ15}.

\subsection{Marginal glass phase}

\begin{figure}
\includegraphics[scale=0.35]{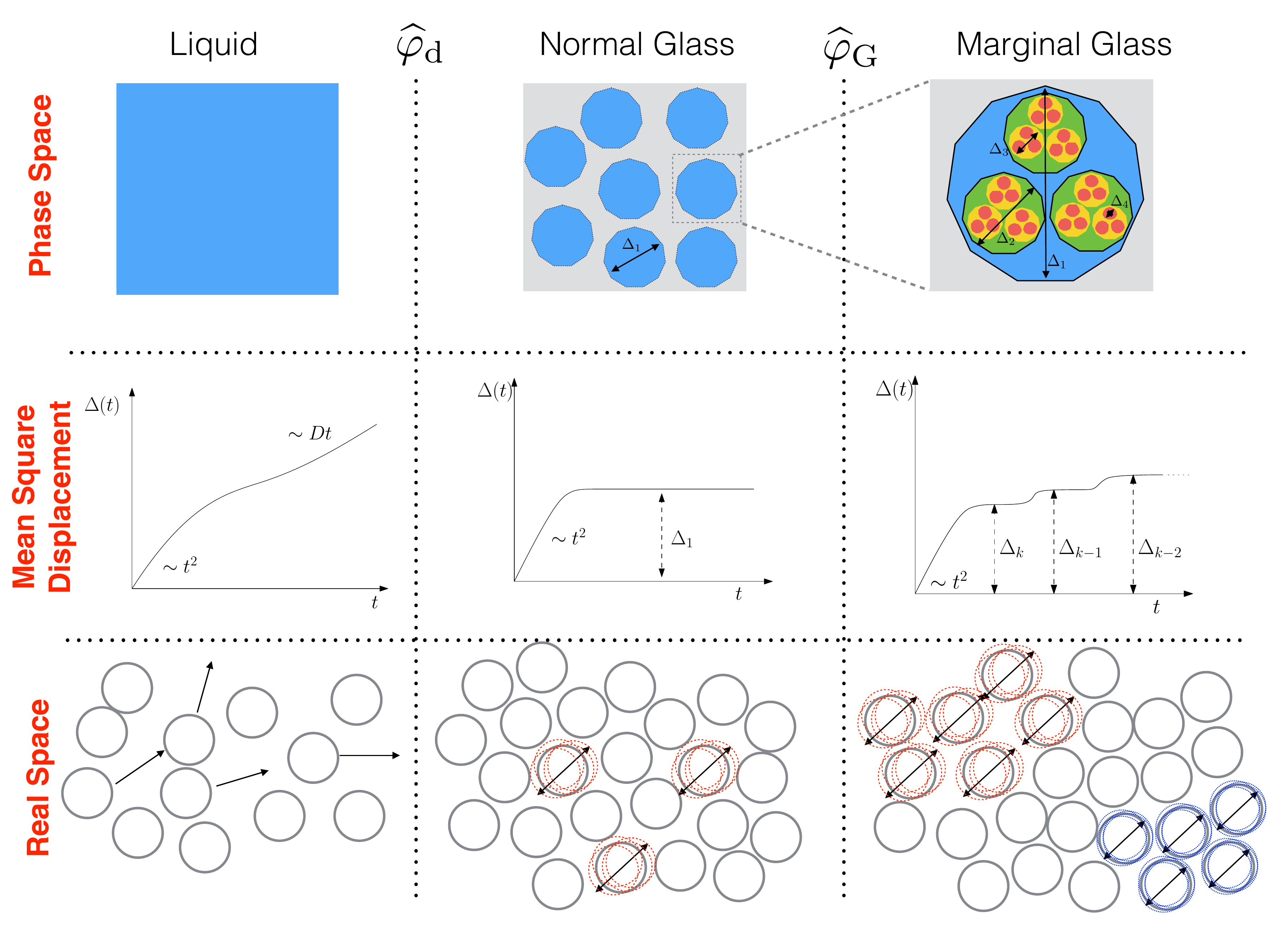}
\caption{The three phases of hard spheres. First column: In the liquid all of phase space can be dynamically reached. The long-time limit of the mean-square displacement is infinite and diffusive. Spheres are not caged.
Second column: For $\wh\f_{\rm g}>\wh\f_{\rm d}$, and for
$\wh \f\in [\wh \f_{\mathrm g}, \wh \f_{\mathrm G}(\wh \f_{\rm g})]$ the system is a normal glass. Phase space is broken into an exponential number of metastables states, which results in disconnected clusters of configurations. Starting from a configuration within a given cluster, only configurations within that cluster can be dynamically reached. Because these configurations correspond to caged spheres vibrating around an amorphous lattice, the mean-square displacement has a finite long-time limit $\Delta(t)\to \Delta_1$ that is proportional to the amplitude of these vibrations. 
Third column: For $\wh \f>\wh \f_{\mathrm G}(\wh \f_{\rm g})$ the system is a marginal glass.
The metastable states become basins of hierarchically organized configurations. Different typical configurations can be at different mutual distance, $\Delta_i$. The long-time dynamics of $\Delta(t)$ is not stationary, but displays an infinite series of plateaus. Vibrations are very spatially heterogeneous with correlated regions of highly vibrating spheres and regions with nearly frozen spheres.
}
\label{fig:Phase_space}
\end{figure}

Beyond the Gardner transition the system enters a marginal glass phase.
Recall that liquid dynamics is diffusive and ergodic (Fig.~\ref{fig:Phase_space} first column), while
in a normal glass there exists an exponential number of metastable states, and dynamically the system quickly and fully explores one basin but cannot visit other basins.
The mean square displacement then has a finite long-time limit $\Delta(t)\to \D \equiv \Delta_1$ (Fig.~\ref{fig:Phase_space} second column).
By contrast, the typical relative mean-square distance between two amorphous lattices corresponding to two different basins is infinite.

At higher densities, however, the system enters the marginal glass phase
(Fig. \ref{fig:Phase_space} third column).
There, various amorphous lattices (glassy states) belong to a same metabasin and are organized hierarchically.
For the sake of illustration, consider a $k$-step replica symmetry breaking
($k$RSB) approximation of phase space in that regime. 
At the very bottom of the hierarchy a state is made by vibrations around an amorphous lattice with an amplitude $\Delta_{k}$.
Two amorphous lattices can be at only $k-1$ different mean square displacements $\{\Delta_{k-1},\ldots,\Delta_1\}$.
For each triplet of amorphous lattices, at least two of the three distances must be equal, which makes the space ultrametric (see Fig.~\ref{fig:Phase_space} top-right panel for $k=4$). The complete solution of hard spheres in $d\rightarrow\infty$ requires
an infinite number of hierarchical levels and $k\to\io$ (fullRSB) with infinitesimally close distances $\D_{i}, \D_{i+1}$.
The spheres vibrate around positions on an amorphous lattice followed by (infinitely) slowly changes of the amorphous lattice itself,
giving rise to a sequence of infinitely close plateaus in the mean square displacement (Fig.~\ref{fig:Phase_space}). 
Moreover, the amplitude of vibrations fluctuates and is correlated over large regions. The system is thus much more heterogeneous than a normal glass.

\subsection{Gardner transition and marginal glass in finite dimensions}

In experiments and simulations, the first challenge to test the above predictions is to obtain fully equilibrated samples at $\wh\f_\mathrm{g}>\wh\f_\mathrm{d}$.  More precisely, equilibrium glass configurations must be obtained for which the characteristic lifetime of the metastable state, $\t_\a$, is much longer than the time needed to equilibrate vibrations within the metastable state, $\t_{\b}$. 
Only in this regime, can an experimental time scale be identified, such that $\t_\a(\wh\f_\mathrm{g})\gg \t_\mathrm{exp}\gg\t_{\b}$. It is therefore possible to
equilibrate inside the metastable glass without escaping it.
Because $\t_\a$ monotonically increase with $\wh \f_\mathrm{g}$, while $\t_\b$ is very large around $\wh \f_{\rm d}$ and decreases for $\wh\f > \wh\f_{\rm d}$,
this condition can only be achieved with deeply equilibrated glasses. In order to distinguish between falling out of equilibrium at $\wh\f_{\rm g}$ and the Gardner transition, one must also reach $\wh\f_\mathrm{g}$ well above $\wh\f_\mathrm{d}$, because the two processes mix and merge for $\wh\f_\mathrm{g}\rightarrow \wh\f_{\mathrm{d}}^+$ (Fig.~\ref{fig:PD_SF}).

Thanks to the existence of activated processes in finite dimension, it is possible to anneal hard spheres beyond $\wh\f_\mathrm{d}$ by compressing the system at a small fixed rate $\RR$, i.e., $\t_\mathrm{exp}=\RR^{-1}$.
In order for the system to trace the equilibrium liquid equation of state, one must keep $\t_\mathrm{exp}\gg \t_\a$ at each moment along the compression. Equilibration becomes impossible when $\t_\mathrm{exp} \sim \tau_\a$, and the system falls out of equilibrium.
Further compression leaves the system stuck inside the last equilibrated metastable glass state at $\wh\f_\mathrm{g}$, as long as $\t_\mathrm{exp}\ll\t_\a$.
And as long as $\t_{\mathrm{exp}}\gg\t_{\b}$ the system remains at equilibrium within the fraction of configuration space restricted to the metastable glassy state in which it got trapped.

Obtaining such fine control over $\t_\mathrm{exp}$ through numerical compression (or thermal annealing in experimental glass formers), however, can be difficult to achieve. The studies that have most carefully examined this regime have thus relied on alternative preparation schemes. In simulations, non-local Monte Carlo sampling has been used to achieve equilibration at densities otherwise unreachable~\cite{BCNO15,BCJPSZ15}; in experiments, vapor deposition has been used to generate ultrastable glasses~\cite{SEP13}. Although the details vary, the key result is that both schemes give access to equilibrium glasses for which $\t_\a\gg\t_\mathrm{exp} \gg \t_\b$.
Such preparation protocol enables the validation of the above theoretical predictions. The most thoroughly documented feature is the hysteresis of glass melting. In fact, the phenomenon was known in physical systems well before a complete theoretical explanation could be provided~\cite{Dy06,SEP13}. The qualitative features of this process are thus fairly robust~\cite{BCJPSZ15}. 

The Gardner transition from the normal to the marginal glass is an original prediction of the $d\rightarrow\infty$ solution.
In finite-$d$ numerical simulations a distinct crossover in glassy dynamics can be seen upon compression, and, as expected, the position of the crossover increases with $\wh\f_\mathrm{g}$ (Fig.~\ref{fig:PD_SF}b). Signatures also appear in the growth of $\t_\b$, $\chi$, and $\xi_\mathrm{G}$~\cite{BCJPSZ15}. Interestingly, the sharpness of that crossover also becomes more pronounced as $\wh\f_\mathrm{g}$ (and thus $\wh\f_\mathrm{G}$) increases. 
Yet, establishing whether the crossover is a true thermodynamic transition and measuring its critical properties
requires a systematic study of larger systems and longer timescales. There is therefore ample room to explore the physics of the Gardner transition both numerically and experimentally~\cite{SD16}. Fuller field-theoretical and renormalization group descriptions might also help grasp the impact of finite-$d$ fluctuations on the transition~\cite{BU15}.


\section{JAMMING}
\label{sec:J}
The endpoint of the compression presented in Sec.~\ref{sec:compression}, i.e., the end of the marginal glass phase,  has a diverging pressure. At that point spheres are in direct mechanical contact with each other; the glass reaches its densest (or close) packing point. 
This point has properties similar to
granular materials at jamming~\cite{LN98,OLLN02,OSLN03,PZ10}, which is
a mechanical rigidity transition~\cite{TS10,BBBCS11} that describes the emergence of solidity in athermal systems, such as foams and grains. These systems jam when an applied pressure leads to slight deformations (elastic or not) of its components. The jamming transition then corresponds to the point where the applied pressure vanishes~\cite{OLLN02,OSLN03}. Theoretical understanding of jamming was further enriched by including marginal stability in the description~\cite{MW15}, as we describe in this section.

\subsection{Mean-field jamming criticality}
As discussed in Sec.~\ref{sec:compression}, the $d\rightarrow\infty$ theory predicts that every equilibrium glass state undergoes a Gardner transition from a normal to a marginal glass at a finite pressure. Hence, the jamming transition, for which $p\rightarrow\infty$, systematically occurs within the marginal glass phase. Like many other glass properties, the density at which jamming takes place, $\wh\f_\mathrm{J}$, depends on $\wh\f_\mathrm{g}$. Moreover, because of the complex state structure in the marginal glass phase, no two compressions ever lead to a same jammed configuration. 

What is especially remarkable is that despite this very strong protocol dependence the critical behavior at and near jamming is universal. For instance, the number of force-bearing sphere-sphere contacts at jamming, $Z$, scales as $Z\sim dN$ in the thermodynamic limit. More impressively, the distribution of weak forces $f$ applied on these contacts as well as the distribution of small gaps $h$ between spheres have a power-law scaling, i.e., 
$P(f)\sim_{f\rightarrow0} f^{\theta_{\mathrm{e}}}$ and $P(h)\sim_{h\rightarrow0}h^{\gamma}$, with irrational exponents $\theta_{\mathrm{e}}=0.42311\ldots$ and $\gamma=0.41269\ldots$, respectively. Part of this robustness likely stems from the hierarchy of subbasins in the marginal glass state upon approaching jamming. As mentioned in Sec.~\ref{sec:compression}, this organization is a fractal with universal dimension $2/\kappa$, where $\kappa=1.41574\ldots$ also describes how the innermost subbasins shrink in size upon approaching jamming, i.e., $\D\sim p^{-\kappa}$. These scalings markedly differ from what one expects in a normal glass or in a crystal~\cite{PZ10,BMSM14}. Indeed, in neither of these cases is a power-law scaling of weak forces or small gaps expected, while $\Delta\sim p^{-1}$ in a normal glass and $\Delta\sim p^{-2}$ in a crystal. 

The exact solution for the vibrational modes around jamming presents a vast excess of low-frequency modes compared with the Debye model for solids (a so-called Boson peak)~\cite{FPUZ15}. The low-frequency limit of the density of states does not vanish as in the Debye model but tends to a constant~\cite{OSLN03,SLN05,WNW05,WSNW05,DLDLW14}. 
The structure of these modes is as extended as phonons, but their organization in no way resembles plane waves~\cite{XVLN10}.

\subsection{Scaling Relations}
Treating jammed states as mechanically marginally stable packings allows for a number of general properties to be worked out independently of the $d\rightarrow\infty$ solution. 
Isostaticity states that a jammed configuration has precisely the number of sphere-sphere force-bearing contacts for the system to be mechanically stable~\cite{Wy12,MW15}.
 More precisely, the saturation of the Maxwell stability criterion gives $Z=dN+\mathcal{O}(1)$, where the correction deterministically depends on the choice of boundary conditions. Note that this result is fully consistent with the scaling of the $d\rightarrow\io$ solution, but more precise. The mechanical marginality analysis further provides relationships for the critical exponents, i.e., $\theta_\mathrm{e} =  1/\gamma-2$ and $\kappa=2/(1+\gamma)$~\cite{Wy12,DLBW14}, 
that are precisely obeyed by the $d\rightarrow\infty$ exponents~\cite{CKPUZ14NatComm,CCPZ15}.
In addition, it predicts the existence of
low-energy localized excitations, which are absent from the exact $d\rightarrow\infty$ solution
and display a different scaling of the weak force distribution with $\theta_{\ell} = 1-2\gamma<\theta_\mathrm{e}$~\cite{Wy12,LDW13,DLBW14,MW15}.

\begin{figure}
\includegraphics[width=\columnwidth]{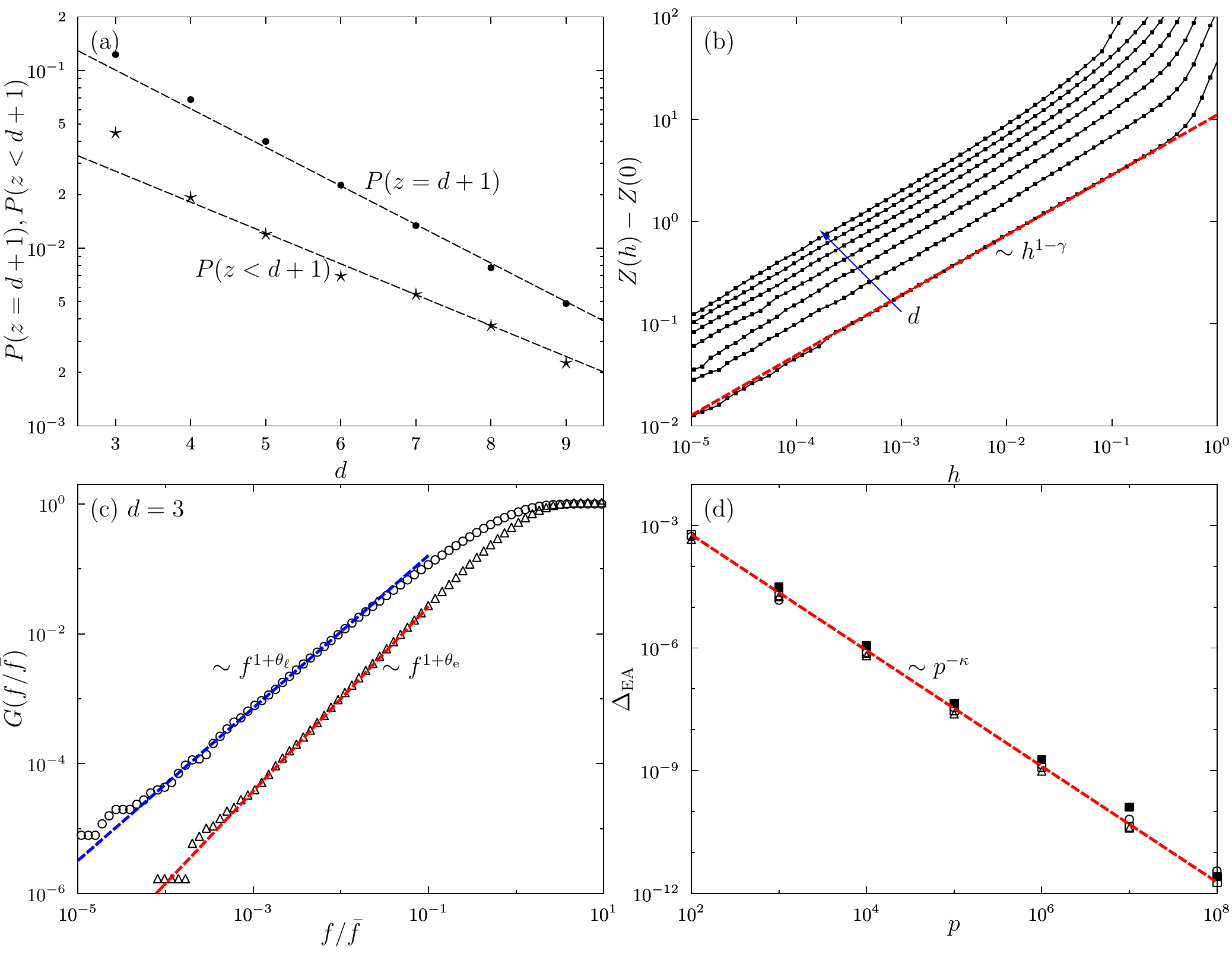}
\caption{
The robust critical behavior around jamming critical is one of the key highlights of the exact solution. It it is, however, only observed when leaving out rattlers (particles with a number of contacts $z<d+1$) and bucklers (with $z=d+1$), (a) whose fractions decay exponentially quickly with dimension~\cite{CCPZ12,CCPZ15}, and are thus nonperturbative effects.  (b) The number of neighbors a distance $h$ from contact, $Z(h)$, then robustly scales as $h^{1-\gamma}$~\cite{CCPZ12}.  (c) Separating the contribution from bucklers (blue) from that of other force bearing contacts (red) reveals the critical scaling of the cumulative distribution function of forces, $G(f)$, for both species~\cite{CCPZ15}. (d) Upon approaching the jamming transition, a nontrivial decay of $\Delta_{\mathrm{EA}}\sim p{-\kappa}$ is robustly observed in $d=3$, 4, 6, and 8 (symbols as in Fig.~\ref{fig:MSD})~\cite{CKPUZ14NatComm}.
}
\label{fig:jamming}
\end{figure}

\subsection{Jamming in finite dimensions}
One of the most remarkable features of the $d\rightarrow\infty$ solution is its agreement with both qualitative and quantitative aspects of jamming observed in numerical simulations. This outcome is especially stunning, because only a few years ago, the notion that $\wh\f_\mathrm{g}$ and $\wh\f_\mathrm{J}$ were actually different transitions was far from generally appreciated.

Since then, a key qualitative aspect of the jamming transition has been unambiguously confirmed. Jamming indeed occurs over a range of $\wh\f_\mathrm{J}$, depending on the preparation protocol~\cite{CBS10,OKIM12,CCPZ12}, and that range increases with $d$~\cite{CIPZ11,CCPZ12}. 
Until now, we have focused on a single preparation protocol to construct jammed states: adiabatic compression of glassy states obtained
from equilibrated liquid configurations at density $\wh\f_{\rm g}$~\cite{PZ10}.
In this specific case, we find that denser equilibrated liquid configurations (higher $\wh\f_{\rm g}$) 
lead to denser final jammed configuration (higher $\wh\f_{\rm J}$). This is clearly observed in the phase diagrams of Fig.~\ref{fig:PD_SF}, both
in the analytical prediction for $d\to\io$ and in the numerical simulation for $d=3$.
We will see in Sec.~\ref{sec:eq_ideal_glass} that the existence of a range of $\wh\f_{\rm J}$ 
remains true when considering more general preparation protocols of jammed states.
It is important to emphasize that this phenomenon is neither a result of the finite-size systems used in simulations,
nor of partial crystallization,
 although both can compound it. 

Many properties of jammed configurations are dimensionally robust and universal:

(i) Particles belonging to the force network at jamming form a perfectly isostatic system~\cite{TS10,GLN12,LDW13,CCPZ15}, thus confirming the prediction, $Z\sim dN$, but in a much more precise way. Only two types of corrections are found, neither having to do with finite-size fluctuations. First, the choice of boundary conditions affects the number of degrees of freedom in the system, and thus a correction of order $d$ must be made to the number of force-bearing contacts. Second, a fraction $f_\mathrm{ratt}$ of particles end up not being part of the force network proper, but rattle within pores of that network. Hence, it is observed that $Z=dN(1-f_\mathrm{ratt}(d))+\mathcal{O}(d)$. Because $f_\mathrm{ratt}$ vanishes exponentially quickly with dimension, however, the exact solution cannot predict this non-perturbative effect (Fig.~\ref{fig:jamming}a). They are thus typically excluded from the structural analysis of jammed configurations.

(ii) It has long been appreciated that the distribution of small interparticle voids in jammed hard sphere configurations displays an anomalous power-law scaling~\cite{TS10}. Numerically, various measures of that exponent give $\gamma=0.40(4)$~\cite{TS10,CCPZ12}. This result holds, irrespective of preparation protocol, for $d=2$ to 12, and agrees with the $d\rightarrow\infty$ prediction (Fig.~\ref{fig:jamming}b).

(iii) Mechanical marginality considerations first predicted that a complementary power-law scaling of the weak contact forces should also be observed~\cite{Wy12}, and an anomalous scaling was noted quickly thereafter in simulations. In order to fully clarify the situation, however, a subtle distinction between particles that are part of the force network is needed. Particles with $d+1$ force-bearing contacts, $d$ of which are nearly coplanar, result in a weak force being associated with the remaining contact. Such particles can buckle in and out of the plane of the coplanar neighbors and give rise to quasi-localized excitations. Because the fraction of bucklers vanishes exponentially with $d$, here again this effect cannot be obtained from the exact solution. But treating bucklers separately from the rest of the force network gives robust critical exponents $\theta_{\mathrm{e}}=0.40(4)$ and $\theta_{\ell}=0.18(2)$~\cite{LDW13,DLBW14,CCPZ15} (Fig.~\ref{fig:jamming}c). 
Here again, these values are in complete agreement with the $d\rightarrow\infty$ solution and with arguments based on mechanical marginality.

(iv) The small excitations associated with opening and breaking force contacts have been separated between extended and quasi-localized excitations~\cite{XVLN10,Wy12,LDW13}. The former is qualitatively predicted from the exact solution, while the latter likely results from the presence of bucklers and of otherwise dimensionally anomalous local geometrical arrangements. A clear geometrical description of these localized excitations remains to be obtained, because these processes are beyond the scope of the $d\rightarrow\infty$ solution.

(v) The anomalous nature of $\kappa$ was suggested from mechanical marginality considerations~\cite{BW09b}, but extracting this quantity is riled with challenges. Unlike the previous exponents, it requires dynamical simulations, and is thus not purely a property of jammed configurations. At finite $p$ within the Gardner phase, all of the processes not captured by the exact solution can further hinder its measurement in low $d$.  Recent numerical determination nonetheless give $\kappa=1.40(4)$ in $d=3-8$~\cite{CKPUZ14NatComm}, which is once again in excellent agreement with the $d\rightarrow\infty$ solution (Fig.~\ref{fig:jamming}d). 

The triangular comparison between the $d\rightarrow\infty$ solution, mechanical marginality considerations, and finite-$d$ numerical results has thus been remarkably productive in clarifying the jamming phenomenology.

\section{SAMPLING GLASSY STATES} \label{sec:eq_ideal_glass}

In the previous sections we have considered what happens to a system of hard spheres that is slowly compressed in the liquid and in the equilibrium glass regimes. 
In this section we take a different approach. We construct a measure that gives equal weight to all existing glassy states at a given density $\wh\f$ and pressure $p$ to determine where
glasses can be found and in what number. This computation allows us to obtain
information on strongly non-equilibrium dynamical protocols for preparing glasses, and is related to the Edwards ensemble.

\subsection{Effective temperature}

We have seen above that at high densities, the equilibrium measure for hard spheres is supported on a large set of 
distinct metastable states, which we now label by an index $\a$.
Our aim here is to sample these states using non-equilibrium weights.
For finite systems, one can construct a partitioning of phase space,
i.e., the unique association of each configuration $X$ to a state $\a$, denoted by $X\in \a$,
by identifying each state with a potential energy minimum $\a$,
and each configuration $X$ to the minimum reached by energy minimization via steepest descent~\cite{SW82,SKT99}.
In the thermodynamic limit, and especially in analytic computations, this procedure cannot be used directly~\cite{BM00,BB04}, but it remains informative -- see, e.g., Appendix A of Ref.~\cite{BJZ11}.
We introduce the free energy of a state and a
generalized partition function~\cite{Mo95}, respectively
\beq\label{eq:falpha}
\eee^{-\b N \mathsf{f}_\a(\wh \f)} = \int_{X\in \a} \de X\eee^{-\b H[X;\wh \f]} \ , \hskip10pt \text{and}  \ \
Z_m=\sum_{\a}\eee^{-\b m N \mathsf{f}_\a(\wh \f)} \ .
\eeq
Here, the control parameter $m$ defines the inverse effective temperature $\beta_\mathrm{eff}\equiv\beta m$ at which states
are sampled.
If $m=1$ the equilibrium partition function is recovered, 
while for $m\neq 1$ states have weights that differ from their equilibrium values.
Defining the configurational entropy $\Si(f,\wh\f)$ then allows us to rewrite the generalized partition function as
\beq\label{eq:Zm_new}
\Sigma(f, \wh \f)\equiv\frac 1N \ln \sum_\a \delta\left[\mathsf{f}-\mathsf{f}_\a(\wh \f)\right] 
\hskip10pt
\Rightarrow
\hskip10pt
Z_m=\int \de \mathsf{f}\, \eee^{N\left[\Sigma(\mathsf{f},\wh \f)-\beta m \mathsf{f}\right]}\:.
\eeq
In the thermodynamic limit, the integral is dominated by states with a free energy $\mathsf{f}^*$ that maximizes the integrand,
i.e., it satisfies $
\left.\frac{\de \Sigma(\mathsf{f},\wh \f)}{\de \mathsf{f}}\right|_{\mathsf{f}=\mathsf{f}^*}=\beta m$.
Because all states that have non-negligible weight in Eq.~(\ref{eq:Zm_new}) have the same free energy $\mathsf{f}_\a(\wh \f) = \mathsf{f}^*$, 
they also have the same pressure $p$. In other words, for fixed $\wh\f$ and $m$ (recall that $\b$ is irrelevant), the 
modified Boltzmann measure mainly contains states with pressure $p(\wh\f, m)$. 
Inverting the relation to fix $\wh\f$ and $p$ as control parameters determines the effective temperature
$m(\wh\f, p)$ 
needed to achieve the desired pressure. 
In the thermodynamic limit the measure used to compute $Z_m$ is then essentially
uniform over all glassy states with fixed $\wh\f$ and $p$.
Using the configurational entropy, $\Si(m,\wh\f) = \Si[\mathsf{f}^*(m,\wh\f),\wh\f]$, which counts the
number of glassy states that dominates the uniform measure, and substituting pressure $p$ for $m$, as discussed above, gives $\Si(p,\wh\f) = \Si[m(p,\wh\f),\wh\f]$, i.e., the (logarithm of the) number of glassy states with pressure $p$ and density $\wh\f$.

An explicit calculation of $Z_m$ and of all the derived quantities, including $\Si(p,\wh\f)$, can be performed using the replica method~\cite{Mo95,PZ10,KPZ12,CKPUZ14NatComm,CKPUZ14JSTAT}. 
The resulting phase diagram in Fig.~\ref{fig:Eq_complexity} indicates where $\D$ has a finite solution (white region). Outside of that region, the number of glassy states
is negligible in the thermodynamic limit. 
In the following subsections, we describe
various aspects of this phase diagram in more details.

\subsection{Equilibrium sampling}

\begin{figure}
\includegraphics[width=1\columnwidth]{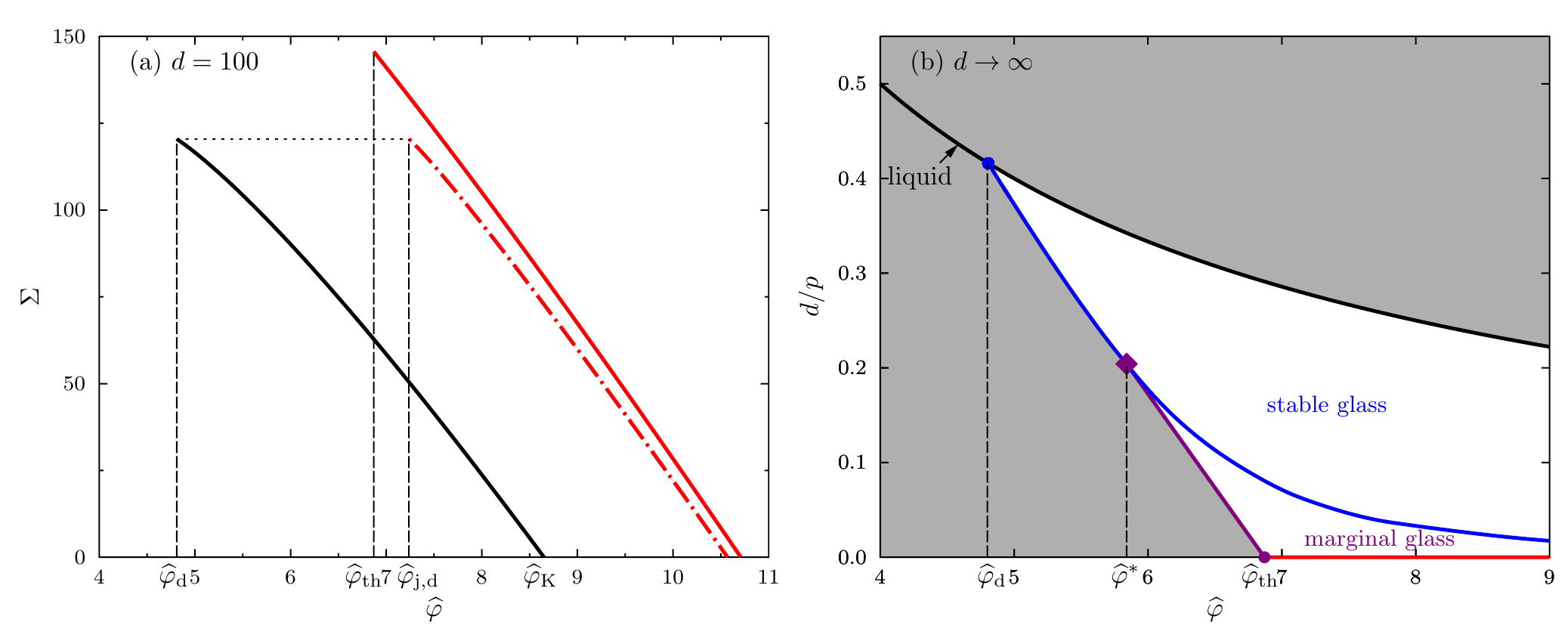}
\caption{(a) Configurational entropy for equilibrium hard spheres at $\wh \f_\mathrm{d}\leq\wh \f\leq\wh \f_\mathrm{K}$ (black), and for jammed states in $d=100$ computed \`a la Edwards (red solid line) and using a slow, adiabatic, compression of the liquid (dot-dashed red line). Adiabatic compression creates exponentially fewer packings than
there exist. 
(b) Phase diagram with generalized Edwards' measure (Monasson real replicas scheme).
The black line depicts the equilibrium liquid line (same as in Fig.~\ref{fig:PD_SF}b);
the blue line depicts the Gardner transition. Between the blue and the violet lines, glassy states become metabasins of marginally stable states. The red line is the jamming line for uniformly-sampled jammed states, i.e., \`a la Edwards.}
\label{fig:Eq_complexity}
\end{figure}

The equilibrium partition function corresponds to $m=1$, where $\b_{\rm eff} = \b$. In this case, we recover the equilibrium line where, for each $\wh\f$,
the pressure of the dominant glass states is the liquid equilibrium pressure $p = p_{\rm liq}(\wh\f)$ 
(black line in Fig.~\ref{fig:Eq_complexity}b). 
We can thus compute the
configurational entropy of equilibrium glassy states, $\Si_{\rm eq}(\wh\f) \equiv \Si[p_{\rm liq}(\wh\f),\wh\f]$, which
is plotted in Fig.~\ref{fig:Eq_complexity}a.
When $\wh \f\geq\wh \f_K\sim \log d$ is reached~\cite{PZ06a,PZ10}, the configurational entropy vanishes.
The ensuing Kauzmann (or ideal glass) transition results from the population of glassy states becoming subextensive in system size. It therefore has a thermodynamic signature, and contrary to the dynamical transition it may remain a phase transition even in finite dimension. Whether that is the case, however, remains an open question. Note that this question is a logically separated issue from 
the existence of metastable states. As such, it is irrelevant for most of the previous discussion about out-of-equilibrium glasses. The relevance
of the $d\to\io$ solution to describe finite-$d$ glasses is not related in any way to the existence of a Kauzmann transition and of an ideal, thermodynamically stable glass.

\subsection{The Gardner line}

For sufficiently high $\wh\f$, it is possible to find states with a pressure, different from that of the equilibrium liquid, $p \neq p_{\rm liq}$.
At each state point in the white zone of Fig.~\ref{fig:Eq_complexity}b we can also compute the stability of the glassy states with respect to the marginal glass phase~\cite{KPUZ13}. The blue line  separates stable from unstable states. In the latter regime, each glass becomes a metabasin of marginally stable glassy states.
This Gardner transition has the same properties as that found in Sec.~\ref{sec:SF_G}, but now distinct groups of glass states are sampled at each
state point $(\wh\f, p)$: in other words, 
moving in the plane $(\wh\f, p)$ does
not correspond to adiabatically following a given group of states. 

\subsection{Jamming line: the Edwards ensemble}

Within the glassy regime, the limit $p\to \io$ is particularly interesting. It corresponds to giving a uniform weight to all stable
jammed states of hard spheres at a given density $\wh\f$~\cite{PZ10}, 
which is precisely the Edwards ensemble prescription. The effective temperature formalism thus represents a generalization of the Edwards measure to finite pressures.
Figure~\ref{fig:Eq_complexity}b shows that there exists a line of jammed states along which the configurational entropy 
$\Sigma_{J}^{(E)}(\wh\f) \equiv \Si[p\to\io,\wh\f]$ can be computed (Fig.~\ref{fig:Eq_complexity}a),
and that this line falls entirely within the marginal glass phase (Fig.~\ref{fig:Eq_complexity}b). Hence, 
all relevant glassy states within the Edwards measure are marginally stable.
As discussed in Sec.~\ref{sec:J} the marginality of glassy states at jamming is responsible for the critical behavior of jammed packings,
and the critical exponents calculated in this ensemble~\cite{CKPUZ14JSTAT} 
are the same as the ones obtained by following
states adiabatically~\cite{RU15}.
This correspondence is quite remarkable. Two very different ways of sampling
jammed packings (uniformly \`a la Edwards, or by slowly annealing the liquid) give the same critical properties. This observation
leads us to conjecture that all jammed states, however sampled, have the same universal critical properties.

This approach also reveals that there is at least one natural algorithm (adiabatic compression) that produces jammed packings
sampled with weights that are distinct from
the Edwards measure.
To prove this point, Fig.~\ref{fig:Eq_complexity} compares the configurational entropy of jammed states produced by
adiabatic compression and of Edwards' jammed states. 
Because for an adiabatic compression the jamming density is a unique function $\wh\f_{\rm J}(\wh\f_{\rm g})$ of the initial equilibrium density,
the configurational entropy of the resulting packings is simply
$\Sigma_J^{(SF)}(\wh \f_{\rm J}) = \Si_{\rm eq}[\wh\f_{\rm g}(\wh\f_{\rm J})]$. This quantity is 
systematically smaller than the Edwards configurational entropy (Fig.~\ref{fig:Eq_complexity}a),
and that this line falls entirely within the marginal glass phase (Fig.~\ref{fig:Eq_complexity}b). Hence, adiabatic compression creates exponentially fewer packings than there exist.
Moreover, the range of $\wh\f_{\rm J}$ that can be constructed via adiabatic compression is smaller than the range of over which jammed packings exist.

\subsection{The threshold and aging dynamics after a crunch}
The phase diagram in Fig.~\ref{fig:Eq_complexity}b shows a third line departing from the equilibrium supercooled liquid line at $\wh \f_\mathrm{d}$.
This threshold line has two branches: one from $(\wh\f_\mathrm{d},p_\mathrm{d})$ to $(\wh\f_*,p_*)$ (purple diamond), where the threshold states are
normal glasses and described by a single order parameter $\D$, and one from $(\wh\f_*,p_*)$ to $(\wh\f_{\rm th}, \io)$ (purple dot), where the threshold states are broken
into a full hierarchy of sub-states. 
Both branches maximize the configurational entropy and correspond to states that are marginally stable. This marginal stability, however, is different than in the Gardner phase. 
Deep in the Gardner phase, far away from the threshold line, metabasins are separated from one another by extensive barriers; it is their interior subbasins that are characterized by a hierarchical structure of marginally stable states.
On the threshold line, by contrast, it is the metabasins themselves that become marginally stable. There are nearly flat directions that connect them with one another, enabling the system to age. Within the metabasins, however, the hierarchical structure of states persists.

In order to understand the physical relevance of threshold states, consider the following experimental protocol.
Take a well-equilibrated liquid at $\wh \f<\wh \f_\mathrm{d}$ and $p<p_\mathrm{d}$, where $p_\mathrm{d}$ is the liquid pressure at the dynamical transition.
The system is then crunched to $p\to p_a>p_\mathrm{d}$, and the time evolution of the packing fraction is monitored.
In the exact solution it is expected that $\wh \f$ increases up to $\wh\f_a$, such that $(\wh \f_a,p_a)$ is on the threshold line.
This strongly out-of-equilibrium protocol is expected to give rise to strong aging.
It can indeed be shown that for large times, if $p_a<p^*$, the response and correlation functions $R(t,t')$ and $\Delta(t,t')$ are related by a modified fluctuation-dissipation relation, where only one effective temperature $\beta_\mathrm{eff}$ defined in Eq.~(\ref{eq:Zm_new}) appears.
For $p_a>p^*$ the situation is still somewhat unclear \cite{Ri13}, but the picture emerging from \cite{CK94,KPUZ13} suggests that aging dynamics is then spread over an infinite number of timescales, each characterized by its own effective temperature. 

\subsection{Out-of-equilibrium glasses in finite dimension}
Studies of finite-$d$, out-of-equilibrium glasses have thus far focused on: (i) developing protocols for generating configurations at the jamming point, and (ii) assessing the validity of the Edwards measure. Although relatively few of the many predictions of the exact solution have been tested, (i) and (ii) establish a strong foundation for exploring these predictions. Because (i) and (ii) have already been extensively reviewed~\cite{TS10,He10,BMOM16}, we here only focus on the aspects most related to the exact solution.

Two main families of out-of-equilibrium protocols have been developed for generating hard-sphere configurations at jamming. The first obeys the volume exclusion of hard spheres throughout the preparation. It thus reaches the jamming transition from densities below it. Typical protocols include the Lubachevsky-Stillinger algorithm, which grows the diameter of hard particles at a fixed rate while running an event-driven molecular dynamics simulation~\cite{LS90}, overdamped event-driven algorithm under external forcing~\cite{LDW13b}, and a sequential linear programming algorithm~\cite{TJ10,HST13}, which iteratively approaches jamming by linearizing the optimization of the packing fraction. The second family of protocols allows particles to overlap during the preparation and steadily minimizes the system energy in order to systematically eliminate these overlaps, thus reaching the jamming transition from densities above it~\cite{OLLN02,OSLN03}. Approaches typically vary in the choice of (purely repulsive) interaction potential and on the initialization condition~\cite{CCPZ12}. 

As mentioned above the key structural properties of the resulting configurations are found to be invariant to the choice of preparation protocol. More subtle aspects, however, remain to be theoretically understood. Although the final density of the jammed configuration is expected to depend on the algorithmic details, systematic changes to the concentration of rattlers at jamming, for instance, are less easily rationalized~\cite{TJ10}.

Over the years, various efforts have probed the Edwards measure both in experiments and in simulations, but the results have often been challenging to interpret~(see Ref.~\cite{BMOM16} for a recent review). In particular, it has been difficult to obtain sufficient control over the preparation protocol to  interpret discrepancies. A rather sophisticated (and computationally demanding) set of methodologies has recently been developed to avoid these pitfalls. For instance, a direct computation of the volume of the basins of attractions of jammed configurations, for a given minimization protocol, has been attempted~\cite{XFL11,APF14}. Yet even under these highly constrained conditions, the observed measure does not match that of Edwards~\cite{PF12,APF14}. Although explanations for these discrepancies have been suggested~\cite{Pa15}, their inclusion within the framework of the exact solution remains to be achieved.

\section{ONGOING AND FUTURE DIRECTIONS}
The advances presented above hint at firm ground for completely solving the glass problem. In closing, we provide an overview actively explored questions.

The rheology of hard-sphere glasses is a key part of the $d\rightarrow\infty$ description that remains to be carefully assessed in finite $d$.
The strain deformation of glassy states up to their yielding spinodal has been obtained from an analysis akin to the compression of glass states~\cite{RUYZ15, RU15}. Remarkably, strain leads to dilatancy and is initially elastic, with higher $\wh \f_\mathrm{g}$ glasses being more rigid. These glasses, however, also systematically undergo a Gardner transition before reaching the yielding point, and higher $\wh \f_\mathrm{g}$ glasses display later transitions. This marginal stability should match the onset of system-spanning avalanches. Direct experimental or computational validation of these predictions, however, have yet to be obtained. 

Beyond the hard-sphere idealization, important avenues of research have yet to be fully explored. Most of the predictions discussed above for hard spheres are thought to apply to a broad selection of liquids, whatever the interaction type~\cite{KMZ16}. The behavior of low-temperature glasses, however, may not be quite as universal. In particular, not all glasses may display a marginal phase. For instance, although the vibrational spectrum of soft spheres is remarkably robust with dimension, and marginality persists even when compressed far above the jamming transition, this regime is eventually extinguished~\cite{CCPPZ15}. 

Open theoretical questions also remain. First and foremost, the dimensional robustness of the jamming description is both remarkable and puzzling. The physical origin of this effect has no fully satisfying explanation. The universality of the phenomenon is also thought to apply far beyond structural glasses~\cite{FP16,FPUZ15}. Second, renormalization group analysis of these effects have yet to be fully developed~\cite{BU15}. Whether perturbative or non-perturbative in nature, finite-$d$ analysis of the different transitions remains a work in progress. Third, the hard-sphere solution proposes tighter packing bounds than known, rigorous results~\cite{CS99} (see~\cite{SZ13} for a recent discussion). 
Formalizing the results, as has been done for the solution of many problems with disorder over the last couple of decades, could thus open a new chapter in the venerable field of discrete geometry.

\section*{DISCLOSURE STATEMENT}
The authors are not aware of any affiliations, memberships, funding, or financial holdings that might be perceived as affecting the objectivity of this review. 

\section*{ACKNOWLEDGMENTS}
The authors acknowledge funding from the Simons Foundation for the collaborative program ``Cracking the Glass Problem''. PC acknowledges support from the National Science Foundation Grant no. NSF DMR-1055586. PU acknowledges support from NPRGGLASS.

\bibliographystyle{ar-style4}
\bibliography{HS}

\end{document}